\documentclass[]{amsart}
\usepackage[foot]{amsaddr}
\usepackage[a4paper,top=2.5cm,bottom=3cm,inner=3cm,outer=3cm]{geometry}

\usepackage[utf8]{inputenc}

\usepackage{amsmath,amsthm}
\usepackage{amsfonts,amssymb,mathrsfs}

\usepackage{color}
\usepackage[svgnames]{xcolor}
\definecolor{myurlcolor}{rgb}{0.6,0,0}
\definecolor{mycitecolor}{rgb}{0,0,0.9}
\definecolor{myrefcolor}{rgb}{0,0,0.9}
\usepackage[bookmarks=false]{hyperref}
\hypersetup{colorlinks,
linkcolor=myrefcolor,
citecolor=mycitecolor,
urlcolor=myurlcolor}

\usepackage[all,2cell,cmtip]{xy}
\UseTwocells
\usepackage{tikz}
\usetikzlibrary{backgrounds,circuits,circuits.ee.IEC,shapes,fit,matrix}

\tikzstyle{simple}=[-,line width=2.000]
\tikzstyle{arrow}=[-,postaction={decorate},decoration={markings,mark=at position .5 with {\arrow{>}}},line width=1.100]
\pgfdeclarelayer{edgelayer}
\pgfdeclarelayer{nodelayer}
\pgfsetlayers{edgelayer,nodelayer,main}

\tikzstyle{none}=[inner sep=-1pt]
\tikzstyle{empty}=[circle,fill=none, draw=none]
\tikzstyle{dot}=[circle,fill=black,draw=black, scale=.4]
\tikzstyle{bounding}=[circle,dashed, fill=none,draw=black, scale=9.00]
\tikzstyle{simple}=[-,draw=black,line width=1.000]
\tikzstyle{inarrow}=[->, >=stealth, shorten >=.03cm,line width=0.6]

\begin{document}

\title{Operads for Designing Systems of Systems}

\author[Baez]{John C.\ Baez$^{1}$} 
\address{$^{1}$Department of Mathematics, University of California, Riverside CA, 92521, USA}

\author[Foley]{John D.\ Foley$^{2}$}
\address{$^2$Metron, Inc., 1818 Library St., Suite 600, Reston, VA 20190, USA}

\email{baez@math.ucr.edu, foley@metsci.com}


\maketitle

\vspace{-0.5cm}

``System of systems'' engineering seeks to analyze, design and deploy collections of systems that together can flexibly address an array of complex tasks.  For example, a smart phone is a system of systems: it includes a touch screen, camera, CPU, internet connection, etc.\ and supports a wide array of applications perhaps not foreseen, but nonetheless enabled, by its original design.  To unlock this flexibility, the Defense Advanced Research Projects Agency launched the Complex Adaptive System Composition and Design Environment (CASCADE) program to develop mathematical foundations for composing and designing complex adaptive systems and demonstrate these foundations in concrete application domains.

We initially anticipated applying existing mathematics for our CASCADE work: specifically, operads and their algebras.  Analogous to how a group abstracts the symmetries of an object, an operad abstracts operations that compose many objects into a single object.  Each ``type" in an operad defines a kind of thing it is able to compose; each ``operation" defines a specific way to compose a number of things of various types into one.  Just as a group can act on a set, an operad can act on something called an ``operad algebra".   When an operad acts on an algebra of this sort, each type is mapped to an actual set of elements of that type, while each operation is mapped to an actual function.  This gives a flexible, general method for building systems of systems from components: operads give the syntax, while their algebras capture their intended meaning, or semantics, in some concrete situation.
Though this general approach to systems has been applied before \cite{NIST, OpWire, CatSci, OpWireYau}, one of our very first use cases required new mathematics.

 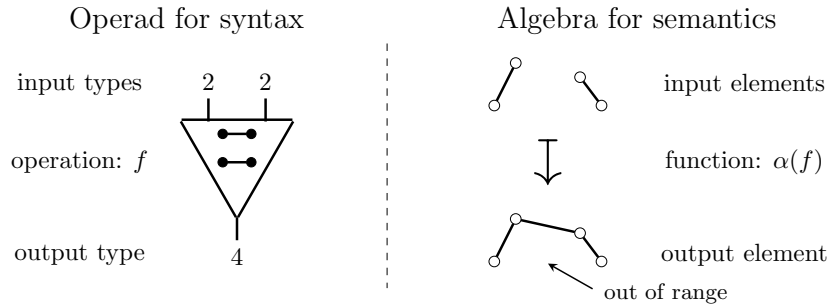
\begin{figure}[h!]
 \centering 
  \[
\scalebox{0.94}{
\begin{tikzpicture}[mydot/.style={
    circle,
    fill=white,
    draw,
    outer sep=0pt,
    inner sep=1.5pt
  }]
	\begin{pgfonlayer}{nodelayer}
		\node [style=dot] (1) at (2, 1) {};
		\node [style=dot] (2) at (2.4, 1) {};
		\node [style=dot] (3) at (2, 0.6) {};
		\node [style=dot] (4) at (2.4, 0.6) {};
		\node [style=none] (a) at (1.4,1.2) {};
		\node [style=none] (b) at (3.0,1.2) {};
		\node [style=none] (c) at (2.2,-.2) {};
		\node [style=none] (c') at (2.2,-.15) {};
		\node [style=none] (D) at (1.8, 1.15) {};
		\node [style=none] (E) at (2.6, 1.15) {};
		\node [above] (D') at (1.8, 1.5) {$2$};
		\node [above] (E') at (2.6, 1.5) {$2$};
		\node [below] (C) at (2.2, -0.5) {$4$};
		\node [style=none] at (1.5, 2.6) {\Large Operad for syntax};
		\node [style=none] at (0, 1.7) {input types};
		\node [style=none] at (0, 0.6) {operation: $f$};
		\node [style=none] at (0, -0.7) {output type};
		
		\node [style=none] (0) at (4.3, 2.3) {};
		\node [style=none] (0') at (4.3, -1.2) {};
		
		\node [style=mydot] (1') at (6.1, 2.0) {};
		\node [style=mydot] (2') at (5.8, 1.4) {};
		\node [style=mydot] (3') at (7, 1.8) {};
		\node [style=mydot] (4') at (7.3, 1.4) {};
		
		\node [style=none] (x') at (8.2, -1.25) {\small out of range};
		\node [rotate=-90, style={font=\Huge}
		] at (6.5, 0.6) {$\mapsto$};
		
		\node [style=mydot] (1'') at (6.1, -0.2) {};
		\node [style=mydot] (2'') at (5.8, -0.8) {};
		\node [style=mydot] (3'') at (7, -0.4) {};
		\node [style=mydot] (4'') at (7.3, -0.8) {};
		\node [style=none] (x) at (7.2, -1.2) {};
		\node [style=none] (y) at (6.5, -0.8) {};

		\node [style=none] at (7.8, 2.6) {\Large Algebra for semantics};
 		\node [style=none] at (9.3, 1.7) {input elements};
 		\node [style=none] at (9.3, 0.6) {function: $\alpha(f)$};
 		\node [style=none] at (9.3, -0.7) {output element};
		
	\end{pgfonlayer}
	\begin{pgfonlayer}{edgelayer}
		\draw [style=simple] (1) to (2);
		\draw [style=simple] (3) to (4);
		\draw [style=simple] (a) to (b);
		\draw [style=simple] (a) to (c);
		\draw [style=simple] (b) to (c);
		\draw [style=simple] (D) to (D');
		\draw [style=simple] (E) to (E');
		\draw [style=simple] (c') to (C);
		\draw [style=simple] (1') to (2');
		\draw [style=simple] (3') to (4');
		\draw [style=simple] (1'') to (2'');
		\draw [style=simple] (3'') to (4'');
		\draw [style=simple] (1'') to (3'');
		\draw [style=inarrow] (x) to (y);
		\draw [style=dashed] (0) to (0');
	\end{pgfonlayer}
\end{tikzpicture}
}
\]
 \caption{
The operation $f$ in the operad $O$ attempts to add two edges connecting two graphs; in a certain algebra $\alpha$ of this operad, the concrete function $\alpha(f)$  adds edges only between vertices that are close enough.}
\label{fig:action}
 \end{figure}
 \vspace{-0.05cm}
 
 This early application concerned different types of aircraft employing secure point-to-point communication.  
 Since an operad was not available ``off the shelf"  to design networks with point-to-point links, we constructed one directly.
 Operations in this operad $O$ are graphs, such as $f$ in Fig.\ \ref{fig:action}; types are the number of vertices of a network.   Each operation gives a formal blueprint for combining a collection of input networks into a single output network by adding edges.

To model system composition, this operad acts on algebras whose elements describe specific aircraft at specific locations.  One of the first interesting mathematical challenges was to model \emph{range-limited} communication as an algebra of this operad. 
When an operation $f$ acts on this algebra, each vertex of the graph $f$ represents an aircraft.
Each edge of $f$ represents a potential communication link.  In the algebra, these are instantiated as actual communication links only if the two aircraft involved obey a condition, e.g.\ being within a certain distance of each other.
 
 Based on this example, we developed a general theory \cite{NetworkModels} which lets us efficiently 
 construct
 ``network operads'' and algebras suitable for a wide range of applications.   The operations in a network operad describe ways to combine several networks to form a larger one.  A network operad acting on an algebra provides a starting point to automatically generate and evaluate candidate system-of-system designs.  It lets us explore formally correct blueprints (operations in that operad) that combine basic systems (elements of the algebra of that operad) into a system of systems.
 
 In the CASCADE project, we applied network operads to domains including maritime search and rescue.  For example, one application problem was inspired by the 1979 Fastnet Race and the 1998 Sydney to Hobart Yacht Race, in which severe weather conditions resulted in many damaged vessels distributed over a large area.  Both events were tragic, with 19 and 6 deaths, respectively.  This sort of disaster remains beyond the scale of current search and rescue planning.  
 
 In the problem we dealt with, various larger assets---e.g.\ ships, airplanes, helicopters---could be based at ports and ferry smaller search and rescue units---e.g.\ small boats, quadcopters---to the search area.   Network operads with directed graph operations were customized to describe allowable nestings of assets and ports.  Algorithms explored designs within budget constraints to achieve a high capacity for search---known in the literature as ``search effort''---that could be delivered in a timely manner.  The most effective designs could ferry a large number of cheap search and rescue units quickly to the scene.

Such applications raised questions about whether limitations on the degree of vertices in a graph---e.g.\ limits on how many quadcopters a helicopter can carry---could be directly encoded into an operad.  This is indeed possible, and the relevant theorems were proved by the first author's graduate student Joe Moeller \cite{Moeller}.

Surprisingly, network operads---originally used to design systems of systems---can also be applied to ``task" them: in other words, specify their behavior.  An elegant example of this approach is given in \cite{NMPetri} where ``catalyst" agents enable behavioral options for a system.  For a search and rescue application to recover downed pilots, operations were built up from primitive tasks that coordinate multiple agent types---e.g., airplanes together with helicopters---to a form a coordinated task plan.  A direct translation of primitive tasks to decision variables for a constraint program---originally a mixed integer linear program and later reworked to leverage the scheduling toolkit of the CPLEX optimization software package---provided a flexible approach to automatically design task plans.  Though this direct approach facilitated correct and transparent modeling for complex tasking problems, only small problems---relative to the demands of applications---were computationally tractable. So, more research is needed to develop efficient algorithms for searching in the set of operations of a network operad.

The operad formalism offers new ways to handle changing levels of abstraction in system-of-system design and tasking.  Both design and tasking are aspects of a multi-stage process in which the structure and behavior of a network is specified in increasing detail, starting from a rough outline.  The initial rough description of a network is typically very abstract.  For example, we can describe a network of boats connected by communication channels without specifying the type of boats, their positions, or the type of channels.  As we proceed further in the process of design and then tasking, we can fill in more details and move to a less abstract description.   Each ``level of abstraction'' can be described by an algebra of a network operad.  ``Changing levels of abstraction'' is then described by a homomorphism between operads, or a homomorphism between their algebras.  We hope more researchers explore the promise of this methodology.

\subsubsection*{Acknowledgment} This work was supported by the Defense Advanced Research Projects Agency (DARPA) under Contract No. N66001-16-C-4048.

\end{document}